\documentclass[showpacs,preprintnumbers,amsmath,amssymb]{revtex4}

\usepackage[dvips]{graphicx} 
\usepackage{dcolumn}
\usepackage{bm}

\begin{document}

\preprint{}

\title{Time structure of muonic showers}

\author{L. Caz\'on$^1$} 
\author{R.A. V\'azquez$^1$} 
\author{A.A. Watson$^2$} 
\author{E. Zas$^1$}
\affiliation{$^1$ Departamento de F\'\i sica de Part\'\i culas, Facultade
de F\'\i sica,\\ Universidade de Santiago de Compostela, 15706
Santiago, SPAIN}
\affiliation{$^2$ Dept. of Physics and Astronomy, University of Leeds, Leeds
LS2 9JT, UK }

\begin{abstract}
An analytical description of the time structure of the pulses induced
by muons in air showers at ground level is deduced assuming the
production distance distribution for the muons can be obtained
elsewhere.  The results of this description are compared against those
obtained from simulated showers using AIRES. Major contributions to
muon time delays are identified and a relation between the time structure
and the depth distribution is unveiled.
\end{abstract}

\maketitle

\section{Introduction}

One of the highest priorities in Astroparticle Physics has become the
understanding of the origin, composition and energy spectrum of Ultra
High Energy Cosmic Rays.  Cosmic rays have been observed for a long
time with energies known to exceed $10^{20}~$eV \cite{NW} but the
remarkably low flux at these energies makes their detection
difficult. The air showers developing in the atmosphere from the
interaction of these particles can be sampled at ground level with
arrays of particle detectors.  Alternatively the fluorescence light
emitted by the nitrogen in the atmosphere can be collected by optical
systems \cite{FE}.  Both types of experiments have confirmed the
existence of cosmic rays of such high energies although there are
discrepancies between the cosmic ray fluxes inferred by the different
techniques \cite{AGASA,HiRes}.

The study of extensive air showers of the highest energies is
difficult because the atmosphere is part of the detector and the
interpretation of measurements is very indirect. The air shower is a
complex chain of interactions resulting in a very large number of
particles from which only sparse data samples are taken. From this
sample, the properties of the primary particle must be deduced.  We
must rely on simplified models, or on complex numerical simulations,
to infer these properties.  These simulations have intrinsic and
unavoidable uncertainties arising from interaction models that require
untested extrapolations of low energy accelerator data.  One of the
complications arises because it is often very difficult to discern
physical effects due to shower fluctuations or to primary composition
from effects due to uncertainties in the interaction
models. It is hoped that through redundancy in the observations, both
the high energy models and the primary particle properties can be
constrained.

Currently, the most important step in this direction is the
development of the Auger Observatory.
The southern part of the Observatory is now under construction 
in Argentina and will
combine the fluorescence technique with a particle array of water \v
Cerenkov tanks covering a surface area of 3000~km$^2$
\cite{Auger}. The signal from the tanks is recorded with Flash Analog to
Digital Converters and relative timing is synchronized using GPS
technology to 10-20~ns. The time structure of the shower front is of
crucial importance because the shower arrival direction in an array of
particle detectors is determined from the relative arrival times of
the signals. This is relatively easy since the shower front can be
approximated by a plane but clearly there are curvature corrections to
the front and the time structure of the signal and its fluctuations
will have an impact on the precision of the reconstruction
procedure. Understanding the time structure of the signal is most
important for controlling the angular resolution. Last but not least, 
preliminary observations at the Auger Observatory are already
revealing that the time
structure of the tank signals carries much information \cite{ICRC} and
it is hoped that this new information can be used to answer
fundamental questions such as primary composition.

As most high energy cosmic ray detectors are, or have been, arrays of
particle detectors, the time structure of the shower front has been
studied before 
\cite{Lapikens,watson,honda,battistoni,Kascade,time1}. 
However most work is either experimental or phenomenological 
and there is not a great deal of knowledge about the
time structure of signals in air showers from the theoretical side or
of its relation to the shower parameters. In this work we study the time
structure of the muon signal in air showers. Our work is based on
ideas that have been introduced for the description of muon densities
in inclined showers \cite{ModelPaper} and is a natural extension of
these ideas. Similar ideas have been developed in \cite{time1,time2,time3}.

Muons are naturally quite energetic when they reach the ground:
otherwise they would have decayed in flight.  As they have long
radiation lengths they do not suffer many interactions in their travel
path from the point where they are created to ground level. As a
result if muons are present in the signal they define the
earlier part of the shower front, i.e. its onset which is usually
chosen to define the arrival time to be used in the reconstruction of
the direction. Also muons have a much flatter lateral distribution
than photons or electrons and they always dominate the signal at
ground level for sufficiently large distances from the shower
axis. Moreover, for inclined showers, they dominate the signal
everywhere at ground level because the electromagnetic part of the
shower, generated by the decay of neutral pions, is absorbed high in 
the atmosphere.  Therefore, the description
of the time structure of muons in air showers is of important
practical interest for the reconstruction of air showers from
measurements made in extensive air shower arrays.  Understanding the
muon part of the shower is not the complete story because there are
important contributions to the shower signal due to electrons and
photons. We will leave the study of the time structure of
electrons and photons for future work.  In this article we neglect the
effect of the earth's magnetic field, although it is known that
magnetic effects are important for the description of muon densities
at high zenith angles. We have checked that the time structure of
muons is not affected by the magnetic field up to zenith angles of 
80 degrees and more, depending on the strength of the magnetic field.
Although our results are general, and we expect that they will
be valid at all zenith angles, we will concentrate on the large zenith
angles range, where the muon signal is more important.
 
The article is organized as follows. In Section II we present the main
assumption on which this work is based and we develop an analytical
description of muon energy distributions in air showers as a function
of transverse distance. These will be used in the next section when
the time structure of the shower front is addressed. In Section III we
describe the two main sources of time delays associated with muons and
we present a model for their description, discussing how it compares
with the results obtained directly using simulations and the relevance
of the magnetic field effects. In Section IV we apply the method to give 
general results about time distributions in air showers and in Section V 
we summarize and conclude.

\section{Towards an analytical description of muons in air showers}

\subsection{Factorization of Muon Distributions}

Our starting point will be an approximate model for the description of
the production of muons in air showers.  As the shower develops, muons
are in general produced at a range of altitudes with energies spanning
a wide range and carrying some transverse momentum.  We will assume
that the muon distribution factorizes completely into a product of
three independent distributions, the energy $f_1(E_i)$, the transverse
momentum $f_2(p_t)$, and the distance to ground level $h(z)$
\begin{equation}
\frac{d^3 N_0}{dp_t dE_i dz } = N_0 f_1(E_i) f_2(p_t) h(z),
\label{factorization}
\end{equation}
where $E_i$ and $ p_t$ are the muon energy and transverse momentum at 
the production point, and $z$ is the distance from this point to ground 
level measured along the shower axis. The relation between $z$ and production 
altitude is straightforward but requires taking into account the curvature of 
the Earth for large zenith angle showers. 
The transverse momentum $p_t$ is defined with respect to the shower axis.
Here the three distributions are normalized to 1 and $N_0$ corresponds
to the total number of muons produced in the shower.  
After production we assume that the muons only undergo continuous
energy loss and decay. This is known to be a reasonable approximation when the
magnetic field is ignored.  Hard catastrophic interactions, such as
bremsstrahlung or pair production, are unlikely because the muon
radiation length in air greatly exceeds the whole atmospheric depth
even for the most inclined showers. The angular deviations of the
muons reaching ground level due to multiple elastic scattering are
negligible because the average muon energy at ground level is about
1~GeV for vertical showers and grows rapidly as the zenith angle
increases to over 200~GeV for horizontal showers \cite{ModelPaper}.

\begin{figure}[htb]
\begin{center}
\includegraphics[width=10 cm]{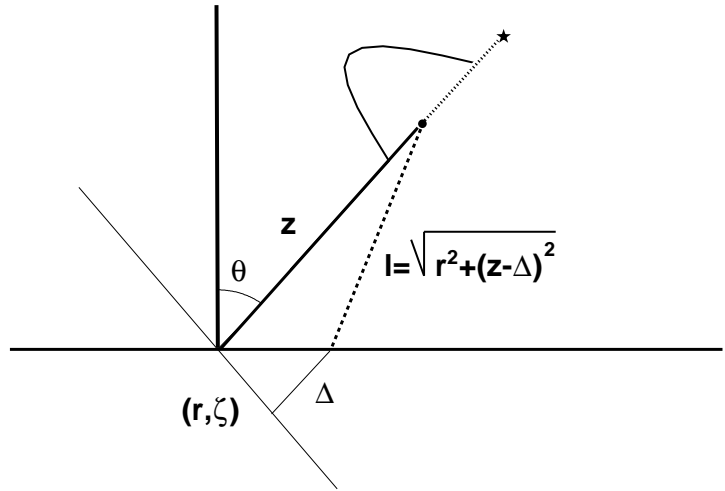}
\caption{Schematic representation of the geometry and notation used
for obtaining the arrival time distribution of muons in air showers.
To describe the position at ground level of the muon signal we will
use a cylindrical coordinate system $(z,r,\zeta)$ with the polar axis
along the shower direction. Coordinates in the transverse plane $r$
and $\zeta$ are defined in such a way that $\zeta=0$ corresponds to the
earliest part of the shower front. The dotted line represents the
distance from first interaction (star) to the muon production point
(dot). The dashed line corresponds to the path traveled by the muon,
$l$, from which the time interval can be deduced. Since muons are
observed as they reach ground level, observation at different angles
$\zeta$ will correspond to different time delays because the path to
ground is shortened by $\Delta$ (which is measured along shower
axis).}
\label{model}
\end{center} 
\end{figure}

In Fig.~\ref{model} we show the geometry of the problem indicating the
notation used.  For a given distance to the shower axis, $r$, the actual
distance traveled by the muon, is simply $l=\sqrt{r^2+(z-\Delta)^2}$,
where $\Delta = r \tan\theta \cos\zeta$. $\theta$ is the shower zenith
angle and
$\zeta$ is the polar angle measured in a plane perpendicular to shower
axis. We choose it to be zero for the case in which $l$ is a minimum.
This system is natural for the description of the asymmetries of the
problem and turns out to be most convenient.

\begin{figure}[htb]
\begin{center}
\includegraphics[width=10 cm]{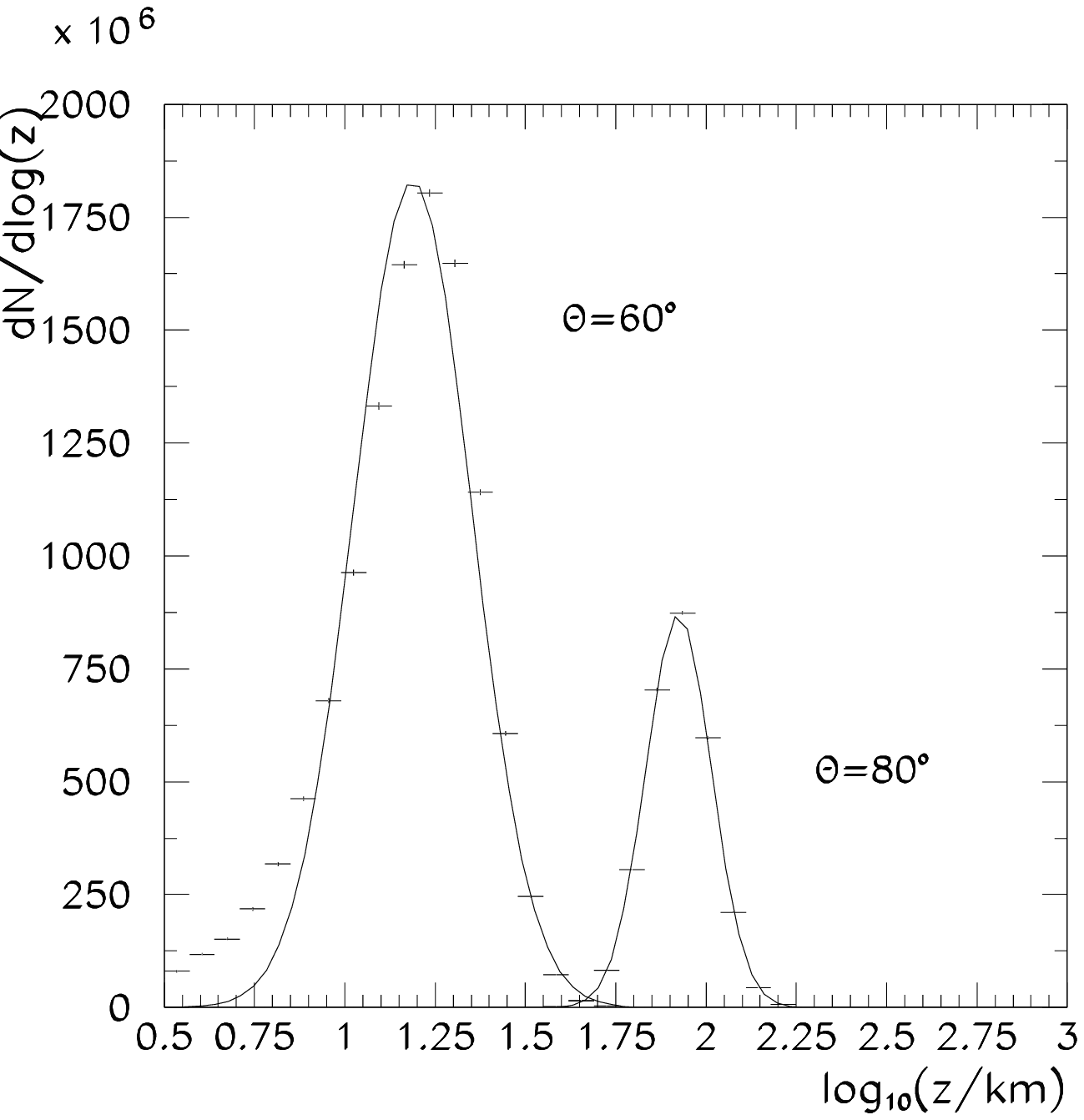}
\caption{Production distance for proton showers of $10^{19}$~eV of 
$60^\circ$ and $80^\circ$ zenith angles as obtained using AIRES 
(histograms).  Also shown is the result of a fit to a log-Gaussian 
distribution (continuous lines). The observation altitude was set to 
1400 m.}
\label{production-dis}
\end{center}
\end{figure}

The distribution of production distance for the muons, $h(z)$, depends
on the details of the hadronic model, on the primary energy and on the
primary chemical composition.  We can parameterize it from Monte Carlo
simulations.  In this article we will use the results obtained with
simulations run with the AIRES code \cite{Aires} for an observation altitude 
of 1400~m, that of the southern Auger observatory, and using the QGSJET model 
for hadronic interactions \cite{QGSJET}. For zenith angles above $60^{\circ}$ 
a Gaussian or a log-Gaussian distribution usually reproduces the production 
profile sufficiently well, but the distributions can be also described 
numerically if necessary. In Fig~\ref{production-dis} we show the
distribution of muon production distance as obtained from AIRES and
our fit to a log-Gaussian distribution.
In Table~\ref{table1} we show the averages ($\log_{10}z_0$) and widths 
($\sigma_z$) of log-Gaussian production distance fits performed for a 
number of showers at different zenith angles.
\begin{table}
\begin{tabular}{|c|c|c|c|}
\hline
$\theta \; \; ({\rm deg.})$ & $z_0$(km)& $\log_{10}(z_0/{\rm km}) $ & $\sigma_z$ \\
\hline
30 & 2.7   &  0.43 &  0.26 \\
50 & 6.6   &  0.82 &  0.21 \\
60 & 13.5  &  1.13 &  0.16 \\
80 & 83.2  &  1.92 &  0.09 \\
\hline
\end{tabular}
\caption{Results of fits to log-Gaussian functions (in decimal base) to the 
production distance distributions obtained with Aires for proton showers and 
an observation altitude of 1400 m for different zenith angles. 
The last two entries are the logarithmic average 
$\log_{10}(z_0/{\rm km})=<\log_{10}(z/{\rm km})> $ 
and the width, $\sigma_z$. 
The second entry is $z_0$ the distance associated to the logarithmic 
average.}  
\label{table1}
\end{table}

\begin{figure}[htb]
\begin{center}
\includegraphics[width=10 cm]{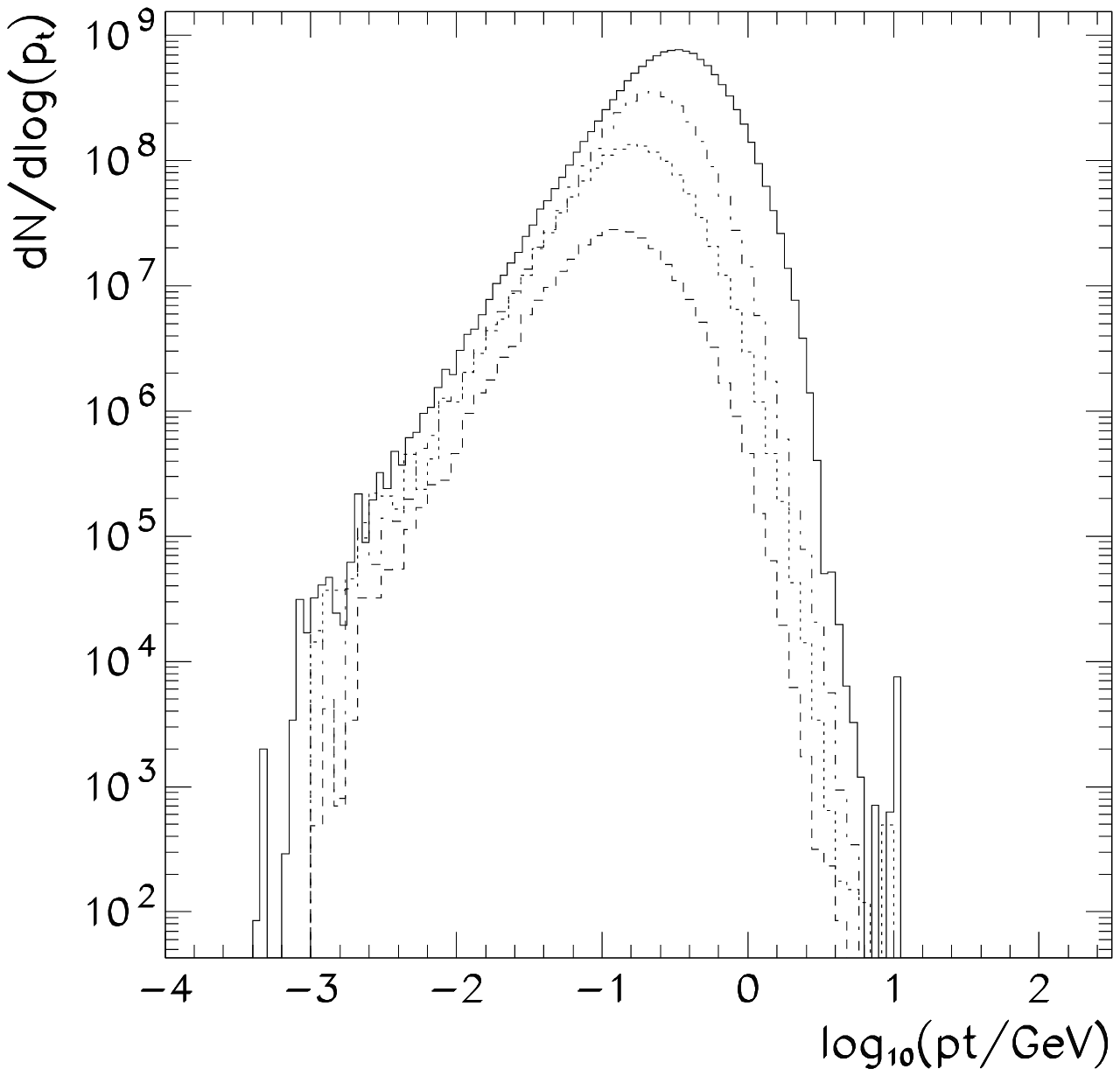}
\caption{Muon transverse momentum distribution at production 
for a proton shower of energy $10^{19}~$eV and $60^\circ$ 
zenith angle as obtained using AIRES. Only muons that reach an
observation altitude of 1400~m are considered. The distribution of all
muons (top curve) is compared to that obtained in different radial
bins, [100,300]~m, [30,100]~m, [10,30]~m (from second top
to bottom).}
\label{pt-dis}
\end{center} 
\end{figure}

Equation \ref{factorization} assumes that the transverse momentum
distribution is independent of both the parent pion energy and its
production altitude. Quite generally, this is equivalent to assuming 
that the transverse momentum distribution is a universal function
which depends only on the hadronic interactions.  When magnetic
effects, muon interactions, and multiple elastic scattering are
neglected, the transverse momentum distribution of the muons at
production is the combination of that of the parent pion and that
introduced through pion decay which is below 30~MeV/c and can be then
ignored when compared to the typical momenta of hadronic interactions
$\sim 200$~MeV/c.  Fig.~\ref{pt-dis} illustrates the transverse momentum
distribution of all the muons that reach ground level. 
The most distinctive feature that can be
observed is a relatively sharp cutoff for large $p_t$. Particularly
for low zenith angles this is close to an exponential cutoff. A good
approximation for $f_2(p_t)$ is given by
\begin{equation}
f_2(p_t) \equiv \frac{1}{N_0} \frac{dN_0}{dp_t} = B p_t^\lambda \exp(-p_t/Q),
\label{pt:distribution}
\end{equation}
where $\lambda \sim 1$, $Q \sim 170$ MeV/c is the characteristic
momentum associated with hadronic interactions, and $B$ is a
normalization constant. 
%

The distribution obtained is actually similar to the transverse
momentum distribution in hadronic interactions as could be expected if
this was the only source of transverse momentum.  The main corrections
to this simplification arise through the cumulative effect of
transverse momenta acquired in multiple scattering of parent hadrons
before the final hadron (usually a pion) decays into a muon.  This
cascading effect contributes to enhance the high-$p_t$ tail of the
momentum distribution, but fortunately it has little effect on the
shape of the bulk of the distribution.  Indeed a small dependence of
the average transverse momentum with production altitude can be
understood, most likely because of this cumulative effect.  As the
width of the $p_t$ distribution greatly exceeds the magnitude of the
shift in transverse momentum due to its correlation with production
distance, one can effectively ignore this dependence which plays a
secondary role.

Finally the energy distribution of the muons is known to be most
affected by muon decay which, depending on the path length traveled by
the muons from the production point to ground level, affects muons of
different energies \cite{ModelPaper,Lorenzo_icrc}. One can, as a first
approximation, use a fixed energy distribution at production,
$f_1(E_i)$, which then evolves because of continuous energy loss
and decay in flight, to give the distribution at ground level, which
is experimentally accessible with an air shower array.  The most
important effect is that there is a strong suppression of the distribution for
muon ground energies below a low energy cutoff which is controlled by
the travel distance for the muons. Clearly this cutoff is heavily dependent on
zenith angle because the average distance to production increases
rapidly from 1~km to 200~km as the zenith angle changes
from vertical to horizontal.

Eq.~\ref{factorization} ignores the transverse position of the parent
particles (mostly pions) that decay into the muons. This is equivalent
to assuming that these particles travel along the shower axis.  A
simple argument shows that indeed pions are confined to a
relatively narrow cylinder compared to the transverse distances of
interest in high energy air shower experiments at ground level. 
A pion of energy $E$ and transverse momentum $p_t$, will
travel at an angle from the shower axis given by $\sin \theta\simeq
c p_t/E$. The pion will on average travel a distance $l = \gamma c
\tau_{\pi}=E/(m_{\pi} c^2) c \tau_{\pi}$ before decaying, which corresponds
to a transverse distance $r_{\pi}=l \sin \theta= \tau_{\pi}
p_t/m_{\pi}$. Given that the $p_t$ distribution is suppressed 
for large $p_t$, we can see that for $\lambda =1$ over $59\%$ of the pions 
will have a transverse distance less than 
$r_{\pi}=  \tau_{\pi} 2Q/m_{\pi} \sim 22$~m because of decay, which can 
be neglected compared to the transverse size of an air shower at ground
level. We also see that $r_\pi$ is independent of the pion energy.  A
simple calculation shows that taking into account the transverse
distance increase due to multiple cascading implies less than a $25 \%$
correction to $r_\pi$
\footnote{In passing, we note that the same argument applies to the 
muon lateral distribution. We then have a characteristic transverse 
scale for the muon lateral distribution that is independent of both 
zenith angle and altitude, $r_\mu \sim \tau_{\mu} 2Q/m_{\mu} \sim$ 
2000 m.}.

Eq.~\ref{factorization} is made simple by ignoring all
correlations between production altitude, energy, and transverse
momentum, but it nevertheless has considerable predictive power.
The fact that these correlations play a subdominant role in what it is
to be described can be understood in terms of the relative importance
of distribution widths with respect to correlation effects.  Although
some correlations are indeed present, we will show that the
assumptions made, which ignore them, predict correctly all the major
characteristics of the time structure of the muon signal in air
showers that we are going to describe.  This can be 
considered sufficient justification to ignore them.

\subsection{Energy distributions at different $r$} 

For the subsequent discussion of arrival time distributions, we will
need details about the energy distributions at fixed transverse
position, $r$, corresponding to typical experimental situations. These
distributions can be reproduced approximately by considering a simple
energy spectrum at production and evolving the distributions accounting for 
energy loss and decay. We will assume $f_1(E_i)$ to be given by a power law
\begin{equation}
f_1(E_i) \equiv \frac{1}{N_0} \frac{dN_0}{dE_i}=A E_i^{-\gamma} 
\Theta(E_i-m c^2),
\label{energy:spectrum}
\end{equation}
where $m c^2$ is the muon rest energy and $\Theta(x)$ is the Heaviside's step 
function. The parameter $\gamma$ can be fitted to Monte Carlo
simulations. Results obtained in such way indicate that $\gamma$ has a
very weak dependence on zenith angle. We have found 
$\gamma=2.6 \pm0.1$ 
to be a reasonable value for all $\theta$.

We assume that the muon energy loss is constant $dE/dx = -k$
where $k=2 $ MeV $/$(g cm$^{-2})$ and take for simplicity a
constant density atmosphere. In this case the muon energy can be
simply expressed in terms of the path traveled, $l$
\begin{equation}
E(l)=E_i-\rho k l,
\label{energyloss}
\end{equation}
where $E_i$ is the muon energy at the production point.  In addition
the number of muons decreases with $l$ due to muon decay. For
a muon with initial energy $E_i$, losing energy in an
uniform density atmosphere, it can be shown that the survival
probability is \cite{Lorenzo_icrc}
\begin{equation}
\frac{N(l)}{N_0} = \left(\frac{E_i - \rho k l}{E_i} \right)^\kappa,
\label{Nmu}
\end{equation}
where the spectral slope $\kappa = m c^2/(c \tau \rho k) \simeq 0.8$ for
$\rho = 1 \times 10^{-3}$ g cm$^{-3}$ \cite{Lorenzo_icrc}. If, instead
of a single muon, we consider a muon energy spectrum at production, 
Eq.~\ref{Nmu} gives the energy spectrum after traveling a distance $l$ 
in terms of the
production spectrum.  It can be seen that there is an effective
low energy cutoff for muons that are produced with less energy than
the corresponding loss during travel time. The low energy behavior of
the ground energy spectrum, here characterized through $\kappa$, is
dependent on the muon energy loss and the atmospheric density. 

It is now straightforward to include energy loss and decay to obtain an
approximate energy spectrum for the muons. We 
take the spectrum at production as given in
Eq.~\ref{energy:spectrum} and multiply it by the
decay probability after traversing a path length $l$. The energy
spectrum of the muons becomes 
\begin{equation}
\frac{dN}{dE_i} = A N_0 E_i^{-\gamma}\left(\frac{E_i-\rho k l 
}{E_i}\right)^{\kappa} \Theta(E_i-m c^2-\rho k l).
\end{equation}
Equivalently, the same distribution in terms of the final muon energy
$E_f$ is given by
\begin{equation}
\frac{dN}{dE_f} = A N_0 (E_f+\rho k l)^{-\gamma}
\left(\frac{E_f}{E_f+\rho k l } \right)^{\kappa}
\Theta(E_f-m c^2).
\label{finalspectrum}
\end{equation}

The energy spectrum discussed so far does not address 
the correlations between transverse distance and 
energy. The factorization assumption 
(Eq.~\ref{factorization}) allows us to explore the $r$-dependence of the 
energy spectra in some detail.  This involves the transverse momentum
distribution which is peaked at $p_t \sim Q$.

According to the assumptions discussed in the introduction, a muon
produced at a distance $z$ from ground level, with energy $E_i$ and
transverse momentum $p_t$ will arrive at ground at a transverse
distance $r$ given by
\begin{equation}
\frac{r}{l}=\frac{p_t c}{\sqrt{E_i^2-(m c^2)^2}}\simeq\frac{c p_t}{E_i},
\label{geometry_relation}
\end{equation}
where $l = \sqrt{(z-\Delta)^2+r^2}$ is the total distance traveled by
the muon and we assume $\sqrt{E_i^2-(m c^2)^2}\simeq E_i$, given that 
$E_i > m c^2+\rho k l$. Eq.~\ref{geometry_relation} relates the transverse
position of the muon to the three independent variables, $E_i$, $p_t$
and $z$ and as a result all that remains is to perform the appropriate
change of variables.

In general we can convert from the $p_t$ distributions to 
$r$ distributions simply by
considering the Jacobian of the transformation 
\begin{equation}
\frac{dp_t}{dr}=\frac{E_i}{c l} \left[1-\frac{r^2}{l^2}\right]. 
\end{equation}
The last term in the above expression can be neglected provided that
$z\gg r$, which is often applicable. 

\begin{figure}[htb]
\begin{center}
\includegraphics[width=10 cm]{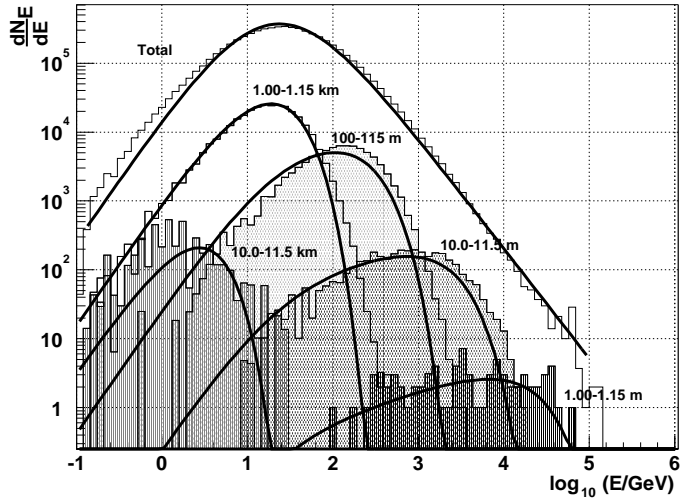}
\caption{Final muon energy spectrum at constant transverse distance, $r$. 
Histograms show Monte
Carlo results obtained with AIRES for $10^{19}$ eV proton showers at
$80^\circ$ zenith angle compared to our prediction for the spectrum
indicated by the continuous lines.  The parameters used for these
results were $\kappa=0.8$, $\gamma=2.6$, $\rho k l =21.1$ GeV.}
\label{check_spectrum_r}
\end{center} 
\end{figure}
If we approximate the $p_t$ distribution by the parameterization given in 
Eq.~\ref{pt:distribution} the spectrum at a given $r$ is given
by
\begin{eqnarray}
\frac{d^2N}{dE_i dr}[z]&=&\frac{A B}{l} N_0 \left(\frac{r}{l}\right)^{\lambda} 
\left[1-\frac{r^2}{l^2}\right] 
E_i^{-\gamma+\lambda+1} \left( \frac{E_i-\rho k l}{E_i} \right)^{\kappa} 
\exp{\left(-\frac{E_i r}{l Q c }\right)} \Theta(E_i-m c^2-\rho k l), 
\label{r:spectrum10} 
\\ 
\frac{d^2N}{dE_f dr}[z]&=&\frac{AB}{l} N_0 \left(\frac{r}{l}\right)^{\lambda} 
\left[1-\frac{r^2}{l^2}\right] \left(E_f+\rho k l \right)^{-\gamma+\lambda+1}
\left(\frac{E_f}{E_f+\rho k l } \right)^{\kappa}
\exp{\left(-\frac{(E_f+\rho k l) r}{l Q c}\right)} \Theta(E_f-m c^2). 
\label{r:spectrum}
\end{eqnarray}
In Fig.~\ref{check_spectrum_r} we show the energy spectrum at constant
transverse distance, $r$, given by the above equation compared to
Monte Carlo results obtained with AIRES for a $80^\circ$ shower.  
We have used the values of $\kappa$ and
$\gamma$ obtained from theoretical considerations and we fit the value
of $\rho k l$ to the overall energy distribution at ground. 
The best fit value corresponds to an energy loss of 
$\rho k l \sim 21.1 $ GeV. 
An alternatively approach could be to fit the parameters 
$\kappa$ and $\gamma$ in
Eq.~\ref{finalspectrum} to the overall energy distribution from the
Monte Carlo data and use them, as effective parameters to predict the
energy distributions at fixed values of $r$. 
The results obtained are not critically dependent on
the specific values of the parameters used nor on the form of the
distributions used as inputs for the calculations. In fact it is well
known that the energy loss parameter $k$ is not constant but increases
logarithmically with the muon energy. By taking it as a fitted 
parameter we include this energy dependence in an effective way.
An analogous procedure can be applied to get the correlations between
other distributions, for instance the $p_t$ distributions for different 
$r$ which are shown in Fig.~\ref{pt-dis}. 

For high zenith angles the production distance distribution is sharply 
peaked and fixing $z$ to $z_0$ is a good approximation. In 
Fig.~\ref{check_spectrum_r} the results have been obtained in this way. 
For low zenith angles this approximation is not justified. 
We can instead consider the
above equations to apply for a given $z$ and integrate them over
the $z$ distribution. For example the energy distribution at fixed $r$ 
becomes then
\begin{equation}
\frac{d^2N}{dE_i dr} = \int \frac{d^2N}{dE_i dr}[z] h(z) dz.
\end{equation}
For close to vertical showers this 
method of performing the $z$ integral reproduces the results obtained
with the simulations with much better accuracy.

A close look at the distributions shown in Fig.~\ref{check_spectrum_r} 
reveals that the energy distributions become broader as the transverse 
distance $r$ decreases. This can be understood in terms of 
Eq.~\ref{r:spectrum}, which predicts in general three distinct regions for
the energy spectrum. 
\begin{enumerate}

\item For $E_f <\rho k l$ the behavior is dominated by the term
corresponding to the muon decay probability, {\it i.e.} by decay and energy
loss. The spectrum is close to a simple power law $dN/dE_f
\sim E_f^{\kappa}$.

\item For $E_f > \rho k l$ the spectrum reflects both the energy and
transverse momentum distributions respectively through the indices
$\gamma$ and $\lambda$. The behavior of the energy spectrum is again
close to a simple power law with a modified index which combines the two
parameters, $dN/dE_f \sim E_f^{-\gamma+\lambda+1}$.

\item For $p_t > Q$ the exponential suppression assumed in the 
transverse momentum distribution appears as a cutoff in 
the energy spectrum. It is simple to show that the suppression takes 
place for energies $E_f > \frac{l Q c}{r}$. 
The cutoff energy grows as $r$ decreases and this makes the 
distributions broader. This suppression affects the average muon 
energy at a given $r$ and is responsible for much of the well-known 
correlation between average energy and transverse position for the muons. 
\end{enumerate}

The three behaviors can be appreciated in the energy distributions for 
small $r$. For large $r$ however the cutoff in transverse momentum 
arises for a final muon energy $E_f < \rho k l$.
These distributions display a direct transition from the $E^{\kappa}_f$ 
behavior to the suppression at the large energy limit. The second
condition ($E_f> \rho k l$) takes place at energies which are suppressed 
and thus has little influence on the bulk of the spectrum. 
We can estimate the approximate distance $r_c$ above which this happens, 
by combining the first and third conditions to obtain 
$r>r_c \sim \displaystyle\frac{Q c}{\rho k} \sim 1000$~m. 
Only for $r$ below this scale the spectrum will display two slopes
discussed above and a cutoff at higher energies. 
It is interesting to notice that $r_c$ is a transverse length scale 
which is approximately independent of zenith angle. It can be seen that 
this scale is also responsible for changes in quantities related to the 
muon energy in showers, for example the correlation between average muon 
energy and transverse distance.

\section{Time delay associated with muon propagation}

The time distribution of the muon signal at ground level is important.
To study this we will consider the delays
associated with different mechanisms in muon propagation.  We 
measure the time delays with respect to a plane front moving in the
shower axis direction and traveling at the speed of light. According
to the simplifications discussed in section II, the muon arrival time
will be determined by a combination of two effects. First there is a
simple geometric effect (geometrical delay), muons arriving at
distance $r$ from the shower axis will take longer, because they
travel a larger path than those arriving at $r=0$. In addition these
muons, being of finite energy, are further delayed with respect to the
reference front moving at the speed of light because they travel more 
slowly (kinematic delay).  We can thus express the total time delay as
a sum of these two delays
\begin{equation}
t = t_g + t_\epsilon.
\end{equation}
Here $t_g$ is the geometrical delay and $t_\epsilon$ is the kinematic
delay.

The distribution of arrival times taking into account both the
geometrical and the kinematic delays can be now obtained by taking a
convolution of the distributions associated with the two effects
\begin{equation}
\frac{1}{N_r} \frac{d^2N}{dtdr}= g(t)\otimes \varepsilon(t)=\int g(t-t')
\varepsilon(t') dt',
\label{convolution}
\end{equation}
where $g(t)$ and $\varepsilon(t)$ are the arrival time distributions
due to the geometrical and kinematical delays respectively, both  
normalized to 1 and $N_r$ is a shorthand notation for $dN/dr$ which is 
needed for this normalization. 
We now study the two delays separately.

\subsection{Geometrical delay}

For high energy muons we can neglect, in a first approximation, the
time delay associated with the finite muon energy. In that case we would
simply convert the path length difference to a corresponding time
delay. For a vertical shower we obtain
\begin{equation}
t_g = \frac{1}{c} \left(\sqrt{z^2 + r^2} -z\right).
\label{t:geometrical}
\end{equation}
This relation implies that the delay of muons arriving at a given
point on the ground, at a fixed transverse distance $r$, is
only a function of the production altitude. The arrival time
distribution of the muon signal can be obtained as
\begin{equation}
g(t) \equiv - \frac{1} {N_r} \frac{d^2N}{dtdr} = 
- \frac{dz}{dt} \frac{1} {N_r} \frac{d^2 N}{dr dz}= 
- h(z) \frac{dz}{dt} \frac{1} {N_r} 
\int dE_i \frac{d^2 N}{dE_i dr}[z],
\label{g(t)-vertical}
\end{equation}
where $h(z)$, $d^2N/dE_idr[z]$ (given in Eq.~\ref{r:spectrum10}) 
and $dz/dt$ are functions of $z$ and therefore 
implicit functions of $t$ through Eq.~\ref{t:geometrical} 
\footnote{Note that $dz/dt$ is negative because the delay decreases as 
$z$ increases. A minus sign has been introduced in the definition of
$g(t)$ to make it positive.}.
It can be shown that the $z$ dependence of the integral of 
Eq.~\ref{r:spectrum10} over $E_i$ is $l^{1-\gamma}~[1-r^2/l^2]$. 
The first factor corresponds to an attenuation of the number of muons through 
decay~\cite{Lorenzo_icrc} and besides the normalization has little importance 
for very inclined showers. 
Moreover since the distance distributions obtained with AIRES only account 
for the muons that reach ground level, it can be argued that the fit takes 
care of it. As a result the functional form of $g(t)$ is given in implicit 
form through
\begin{equation}
g(t) \propto - h(z) \frac{dz}{dt} \left[1-\frac{r^2}{l^2}\right] 
\label{g(t)-shape}
\end{equation}
where for large zenith angles the last factor can be dropped. 

It is relatively easy to understand that for inclined showers the geometrical
effect implies that there is an asymmetry in the arrival times. The
different path lengths traveled by the muons for different polar
angles in the transverse plane ($\zeta$) (see Fig.~\ref{model}) induce
different delays.  We can take this into account by changing the
relation between $z$ and $r$ of Eq.~\ref{t:geometrical} to include the
excess path traveled by muons at different angles
\begin{equation}
t_g(z-\Delta) = \frac{1}{c} 
\left(\sqrt{(z-\Delta)^2 + r^2} -(z-\Delta)\right),
\label{t_g-inclined}
\end{equation}
where $\Delta = r \cos \zeta \tan \theta$, and $t_g$ is measured with
respect to the shower front plane. Now we still use
Eq.~\ref{g(t)-vertical} to get the time distribution but $dz/dt $ must
be obtained from Eq~\ref{t_g-inclined}.  Alternatively we can make the
shift $z \rightarrow z+\Delta$ in the argument of $h(z)$
\begin{equation}
g(t) = -\frac{1}{N_r} \frac{d^2N}{dt dr} 
\propto - h(z+\Delta) \frac{dz}{dt} \left[1-\frac{r^2}{l^2}\right] ,
\label{profile:geometrical}
\end{equation}
and still use Eq.~\ref{t:geometrical} for $dz/dt$.  As in our
convention $z$ is measured along the shower axis from ground to
production point, a given $dz/dt$ from Eq.~\ref{t:geometrical} must be
associated with a production point which is shifted by $\Delta$ to
ensure that distance to production corresponding to a point
$(r,\zeta)$ is precisely $z$.  The above equation introduces, in a
simple way, important asymmetries in time distribution for
constant $r$ but different $\zeta$ angles.

\begin{figure}[htb]
\begin{center}
\includegraphics[width=10 cm]{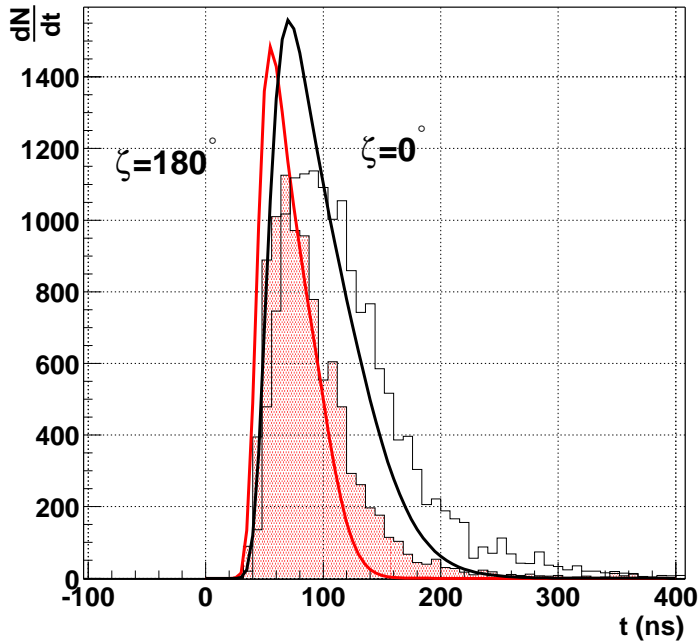}
\caption{Arrival time distribution of muons at $r=$2000 m 
for a $10^{19}~$eV proton shower and incident 
zenith angle of $80^{\circ}$ obtained with AIRES, for two values of 
$\zeta=0, 180^\circ$. The analytical 
result (continuous lines) with which it is compared 
takes into account only the geometrical
delay. The observation altitude is 1400 m.}
\label{h4_4}
\end{center} 
\end{figure}

In Fig.~\ref{h4_4} we show an example of the pulse shape obtained
using the result of Eq.~\ref{profile:geometrical} compared to the
results obtained using Monte Carlo simulation carried out with AIRES.  
The agreement is reasonably good although there are significative
differences. The predicted signal has a faster onset, is somewhat
broader and has a shorter tail which is suggestive of a delay
underestimate.  We will show below that all these effects can be
mostly attributed to the sub--luminal muon velocity.

Once we have an expression for the time distribution we can obtain all
relevant statistical quantities associated with it. For example the
average arrival time at a fixed transverse distance is simply given by
\begin{equation}
<t(r)> = \int_0^\infty dt ~ \; t ~\; g(t) = \int_0^\infty dz ~\;
     h(z+\Delta) \; ~ t_g(z).
\end{equation}

For example, for $r \ll z$, as is the case in
inclined showers, we can expand $t_g$ given by
Eq.~\ref{t:geometrical}, to get a simple expression relating the
time delay and the transverse distance
\begin{equation}<t(r)> = \frac{1}{2c} r^2 \int_0^\infty \frac{dz}{z} h(z+\Delta).
\end{equation}
For $h(z)$ we can use a log-Gaussian distribution centered at $z_0$
with a standard deviation $\sigma_z$, which is a good approximation for
inclined showers (see Fig.~\ref{production-dis})
\begin{equation}
h(z) = \frac{C}{z} \exp \left[-\frac{1}{2 \sigma_z^2}
(\log_{10}(z/z_0))^2\right],
\end{equation}
where $C$ is a normalization constant. When we substitute the above
distribution into the expression for the average geometric time delay
we get the following analytical expression 
\begin{equation}
<t(r)> = \frac{1}{2c} \frac{e^{-\sigma^2/2} \; r^2}{z_0-\Delta}.
\label{t:geom:average}
\end{equation}
where $\sigma=\ln (10)~ \sigma_z$. 
The above expression gives the time curvature as obtained using the
average arrival times of the signals as a function of the mean
production depth. This curvature is directly related to the naive
geometrical radius of curvature given by $1/2~r^2/z_0$.  We notice
that there is a correction factor (involving the width of the muon
production function distribution) between the naive geometrical
curvature and the time curvature.

We can also obtain the RMS value of the arrival time distribution of
the muons at ground level as a function of $r$, $z_0$ and
$\sigma$. The result is
\begin{equation}
\sigma_t(r)  =  \sqrt{<t^2> -<t>^2}  
             =  \frac{1}{2c}~ \frac{r^2}{z_0-\Delta}
\sqrt{1-e^{-\sigma^2}} \sim \frac{1}{2c} \frac{r^2}{z_0-\Delta} ~ \sigma.
\label{sigma-t-r}
\end{equation}
This illustrates that the width of the arrival time distribution is
proportional to the mean value, and hence it also rises as $r^2$ for a
given shower. If $\sigma^2$ is small the proportionality factor
becomes simply $\sigma$ as indicated in the last equality of the above
equation.

In the same way we can obtained an estimation for the asymmetries in
the geometric delay. We will define the asymmetry as 
\begin{equation}
A = <t(r,\zeta=180^\circ)> - <t(r,\zeta=0^\circ)> 
\end{equation}
Then a simple calculation using the distributions introduced before
shows that
\begin{equation}
A = \frac{ r^3 \tan \theta}{c z_0^2} e^{2 \sigma^2}.
\end{equation}
The asymmetry grows rapidly with $r$, as $r^3$. 
Also, notice that at fixed $r$ the average time can be approximated by
a dipole type formula, $<t> = a + b \cos\zeta$. 

\begin{figure}[htb]
\begin{center}
\includegraphics[width=10 cm]{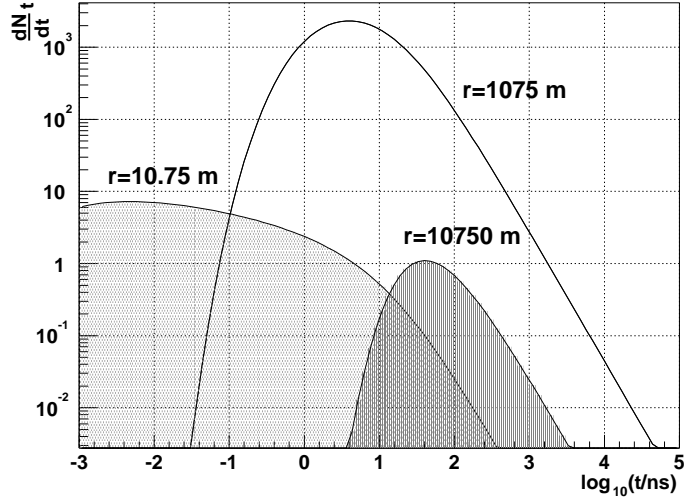}
\caption{Kinematical delay distributions for $r \sim $ 10 m, 1 km, and 10 km 
obtained analytically. The parameters used are
$\kappa=0.8$, $\lambda=1$, $Q=0.17$ GeV/c, and $\gamma=2.6$. }
\label{energy_pulses_log}
\end{center}
\end{figure}

The main features of the arrival time distributions can be described with
this simple model. As obtained in the above expressions, both the time
delay and the width of the distribution grow as the transverse
distance increases.  The asymmetries obtained for different angles
$\zeta$ are in agreement with the simulation results. It is
interesting to notice that the arrival times for small polar angles
($\zeta \sim 0^ \circ$), corresponding to the part of the shower front
that touches ground earliest, are further delayed with respect to the
late arriving part ($\zeta \sim 180^\circ$). We can expect
different curvature corrections for different polar
angles. The asymmetries are largest for intermediate zenith angles,
according to this simple model. In the
actual experimental situation, typically involving an array of
particle detectors, the curvature of the front is sampled at different
angles $\zeta$ so that systematic corrections to the arrival times and
curvature of the front can be expected from this effect. 

The results in Eq.~\ref{sigma-t-r} 
may be useful when considering the weight to be 
given to a given time in the reconstruction of the arrival direction. 
These results can be significantly improved by considering the delay time
due to the muon velocity which we now describe.

A feature can be deduced from Eq.~\ref{profile:geometrical}
which, in principle, opens up a new way to extract relevant
information from the time structure of the shower front.  It implies
that a measurement of the arrival time of muons at a given distance to
the core can give information on the production profile function of
the muons.  Important shower properties such as the
shower maximum, which is related to the maximum in the muon production
function, can in principle be obtained from it, to the extent that
this approximation is valid.

\subsection{Kinematic delay}

There is an additional delay to be considered because muons with finite energy
travel with a smaller velocity than that of light. This introduces
the additional time delay $t_\epsilon$. For instance a 1 GeV muon
traveling 40 km in vacuum would be delayed 750 ns with respect
to the light front. In a medium such as air, muons have continuous energy 
loses and these delays are more complex to estimate.  For
this reason we have modeled the energy distributions in Section II-B,
taking into account energy losses and muon decay.

Neglecting multiple scattering and magnetic effects, muons travel a
distance $l=\sqrt{z'^2+r^2}$ from production to ground level, where
$z'=z-\Delta$.  With these simplifications the extra travel time taken
by a muon of velocity $\beta c$ with respect to a photon is simply
given by
\begin{equation}
t_\epsilon=\frac{1}{c} \int_0^{l}dl' \left[ \frac{1}{\beta(E)}-1
\right].
\label{deltat}
\end{equation}
We can substitute into the above equation the muon energy as a
function of travel distance $l$ and then integrate. This can be done
approximately using Eq.~\ref{energyloss} to obtain
\begin{equation}
t_\epsilon= \frac{1}{c \rho k} 
\left( \sqrt{E_i^2 - (m c^2)^2}-\sqrt{E_f^2 - (m c^2)^2} \right)-\frac{l}{c},
\label{delay_exact}
\end{equation}
where $E_f = E_i -\rho k l $ is the final muon energy when it reaches
the ground.  An interesting feature is revealed by this
equation. It leads to the conclusion that there is a maximum time
delay which is
independent of $l$ and is approximately given by $mc^2/c \rho k\simeq
1.7~\mu$s. It corresponds to a muon that loses all its energy after
traveling the distance $l$.  This upper bound to the kinematic delay
increases slightly if we consider an exponentially decaying density
atmosphere.

If the final energy is much larger than the muon mass, $m c^2 \ll E_f$ we
can expand Eq.~\ref{delay_exact} to obtain
\begin{equation}
t_\epsilon \simeq \frac{1}{2}\frac{(m c^2)^2}{c \rho k }
\left[\frac{1}{E_i-\rho k l} -\frac{1}{E_i}\right].
\label{t:kinetic} 
\end{equation}

When we take into account the muon energy spectrum, for instance in
terms of production energy $E_i$, the arrival time distribution is
then given by
\begin{equation}
\varepsilon(t_\epsilon;z,r)\equiv-\displaystyle \frac{1}{N_r}
\frac{d^2N} {dt_\epsilon dr}=-\frac{1}{N_r} \frac{d^2N}{dE_i
dr}\frac{dE_i}{dt_\epsilon}.
\label{kinetic-distribution}
\end{equation}

At this stage, we make some simplifying assumptions to
obtain simple expressions and discuss qualitative effects. If we
expand Eq.~\ref{t:kinetic} for muon energies that are large in relation to the
energy loss in the atmosphere we obtain the following relation between
time delay and muon energy
\begin{equation}
t_\epsilon \simeq \frac{1}{2}\frac{(m c^2)^2 l}{c E_i^2 }.
\label{t:kin-appr} 
\end{equation}
If we now assume a fixed $p_t$ we can relate $E_i$ to $r$ to get
\begin{equation}
t_\epsilon \simeq \frac{1}{2}\frac{(m c^2)^2}{c p_t^2} \frac{r^2}{z'}
\label{corr-curvataure} 
\end{equation}
where we have expanded $l$ as a function of $z'$ and $r$ for $r\ll z'$.
This is quite interesting because we obtain a kinematic delay which
has a very similar expression to that obtained for the geometric time
delay in Eq.~\ref{t:geom:average}. Very roughly one could expect an
extra correction of order $m^2/p_t^2$ due to the kinematic delay.
Assuming that the $p_t$ distribution is sharply peaked at about
$170$~MeV/c this would result in $\sim 25\%$ more curvature.  When the
geometric delay is compared to simulation results, a systematic effect
of this nature and of similar magnitude is indeed observed as
remarked before. However the approximations involved to obtain this result 
are rather poor.

Fortunately we can obtain more reliable time distributions due to the
kinematic delay through Eq.~\ref{kinetic-distribution}.  We first
invert Eq.~\ref{t:kinetic} to get
\begin{equation}
E_i=\frac{\rho k l}{2}\left[1+ \sqrt{1+ \frac{t_c}{t}}\right],
\label{Eoft}
\end{equation}
Here $t_c$ is a characteristic time that involves energy loss and the
atmospheric density
\begin{equation}
t_c=2\left(\displaystyle \frac{m c^2}{\rho k}\right)^2
\displaystyle\frac{1}{c \, l}\simeq 500 \; {\rm ns} \; \left[ \frac{1
\; {\rm km}}{l} \right].
\end{equation}
We now have the ingredients to calculate the
kinematic time delay distributions. We have only to substitute the
time derivative and the muon energy and transverse distance
distributions (for instance as obtained in Section II) into
Eq.~\ref{kinetic-distribution}.  The final result is cumbersome and we
do not write it down explicitly.

As discussed in Section II-B the energy distributions have different
behaviors depending on the transverse distance $r$ in relation to the
critical radius $r_c$. We can consider the two cases separately 
to better understand the effect of the energy distribution on
the time structure of the pulses.  For $r \gg r_c$ we can simplify the
energy distributions in Eq.~\ref{r:spectrum} to a combination of a
power law and an exponential. In that case the time distribution is
simply
\begin{equation}
\varepsilon(t)\propto t^{-\kappa-2} \exp{\displaystyle
\left(-\frac{1}{2}\frac{(m c^2)^2 r}{ \rho k l c Q} \frac{1}{ct}\right)}.
\end{equation}
For $r \ll r_c$ the time structure is more complex because we obtain two
different behaviors in the two regimes described.  This can be seen
through Eq.~\ref{Eoft} in which the energy is regarded as a function
of the delay time.  The two regimes correspond to the times well below
and well above $t_c$.  These regimes also correspond to the limits
$E_f \gg \rho k l$ and $E_f \ll \rho k l$ respectively.  In these two
limits the expressions for the energy in terms of $t$ can be
simplified to
\begin{itemize}
\item for $t \ll t_c \; \; \; $ $E_i \simeq \sqrt{\displaystyle
\frac{1}{2}\frac{(m c^2)^2}{ct}l}$
\item for $t \gg t_c \; \; \;$ $E_f \simeq \displaystyle
\frac{1}{2}\frac{(m c^2)^2}{\rho k ct}$
\end{itemize}
When these relations are used together with the energy distributions
to obtain the time distribution through Eq.~\ref{kinetic-distribution}
the two limiting behaviors obtained are
\begin{equation}
\varepsilon(t)\propto\left\{ \begin{array}{ll} t^{-\kappa-2} &
\textrm{if $t \gg t_{c}$}\\ \\ t^{\frac{\gamma-\lambda-4}{2}}
\exp{\displaystyle\left(-\frac{1}{\sqrt{2 l c}}\frac{m c^2 r} {c Q}
\frac{1}{\sqrt{t}}\right)} & \textrm{if $t \ll t_{c}$}\\
\end{array} \right.
\end{equation}

Although we have made some effort to give analytical
parameterizations where possible, making several approximations,
we stress here that the complete procedure, without these
approximations, can be implemented. Note that $r_c$ is below the
distance separation between detectors in the Auger Observatory. 

\section{Results} 

\begin{figure}[htb]
\begin{center}
\includegraphics[width=10 cm]{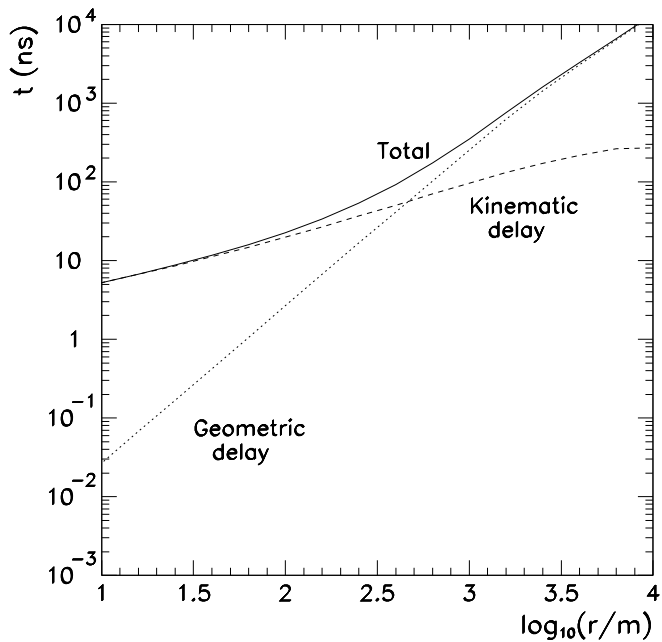}
\caption{Average time delay for muons injected at 10 km with a $\gamma = 2.6$ 
spectrum. The lower curve shows the geometric delay. The dashed points show 
the kinematic delay. The upper curve is the sum of both. }
\label{ave_kg}
\end{center} 
\end{figure}
We can combine the geometric and kinematic delays through the
convolution given in Eq.~\ref{convolution}.  
In Fig.~\ref{ave_kg} we show the average pulse time delay 
for the geometric and the kinematic delay times. We see that at small
distances to the core the delay is dominated by the kinematic
delay. This may seem surprising as near the core the characteristic muon 
energies are larger and one would expect a lower kinematic delay. However,
near the core the spread on energy is larger (see Fig.~\ref{check_spectrum_r}) 
and the time delay is dominated by low energy muons. 

\begin{figure}[htb]
\begin{center}
\includegraphics[width=10 cm]{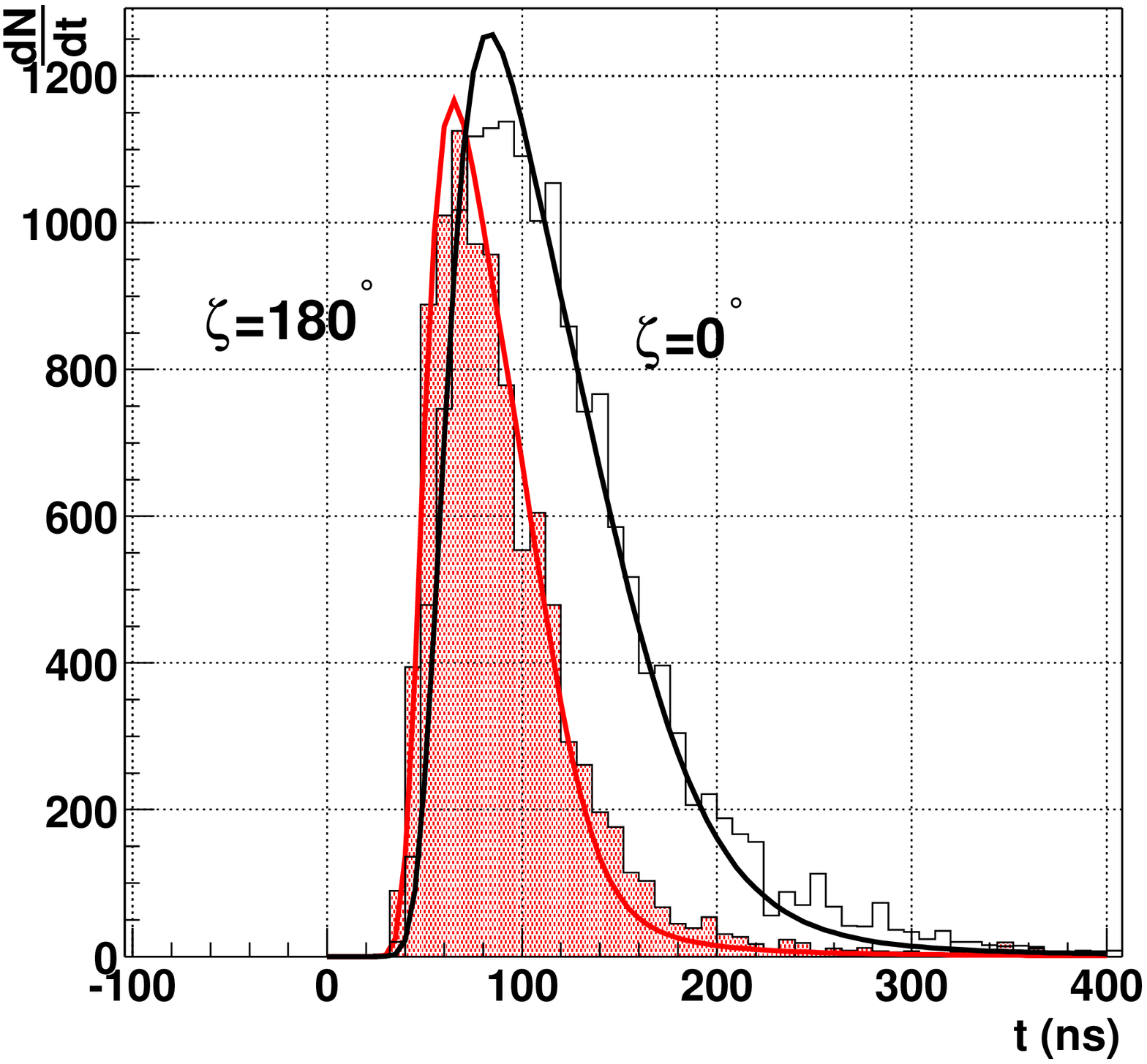}
\caption{Example of arrival time distribution from Monte 
Carlo (histograms) for a $10^{19}$ eV proton shower and 80$^\circ$ 
zenith angle compared to our analytical calculation. Muons arriving 
in the range  $3.2< \log_{10} r<3.4$ ($r\sim $ 2000 m) are selected
for two $\zeta$ angles in the transverse plane.}
\label{h4_5}
\end{center} 
\end{figure}
\begin{figure}[htb]
\begin{center}
\includegraphics[width=10 cm]{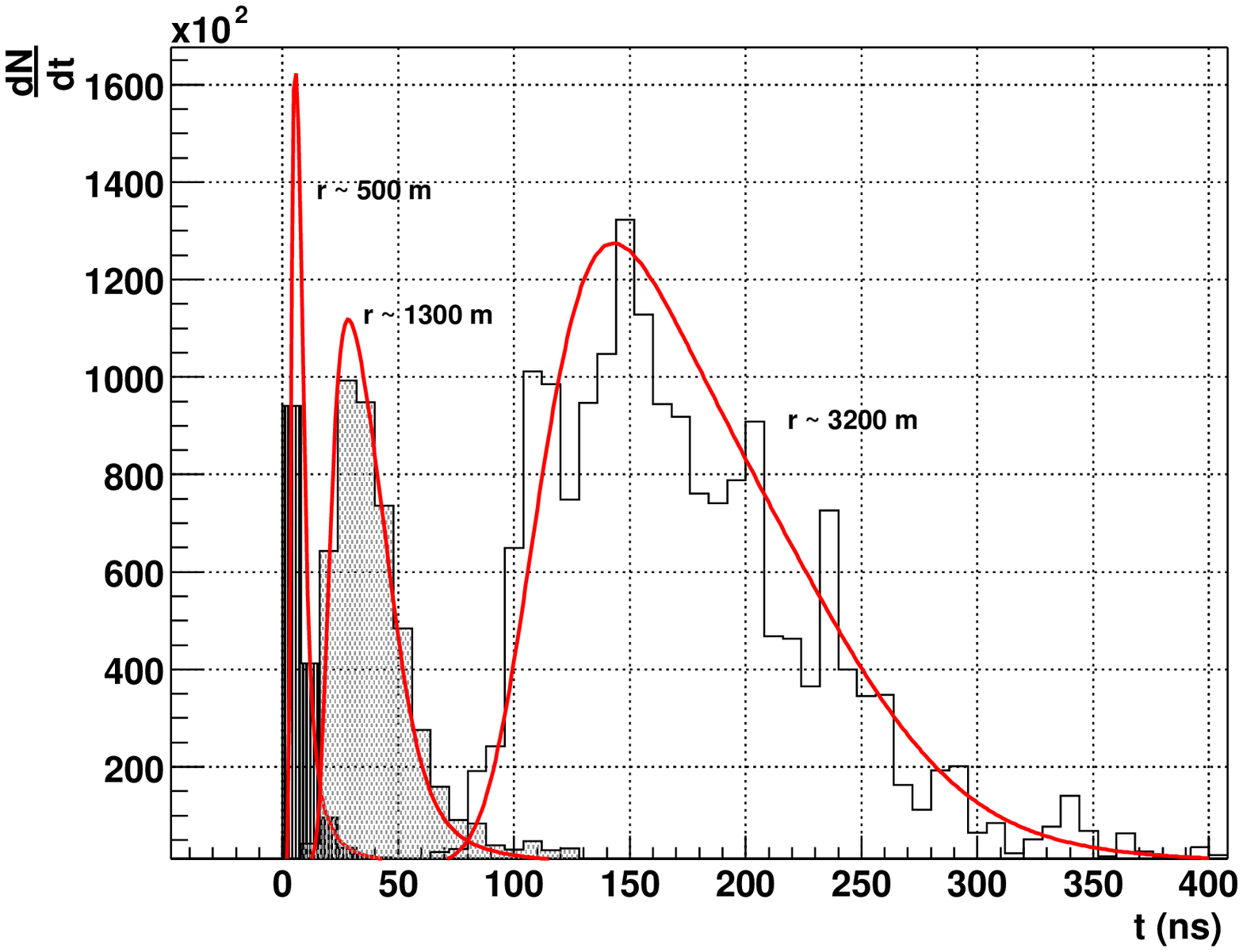}
\caption{Example of arrival time distribution from Monte Carlo 
(histograms) for a $10^{19}$ eV proton shower and 60$^\circ$
zenith angle and $\zeta = 180^\circ$ compared to our analytical calculation. 
Muons arriving in the range  $2.6< \log_{10} r<2.8$, $3.0< \log_{10} r<3.2$, 
$3.4< \log_{10} r<3.6$ as marked.} 
\label{h4_6}
\end{center} 
\end{figure}
In Fig.~\ref{h4_5} we show our analytical result for 
the arrival time distribution compared to the Monte Carlo. 
We can see that the agreement is very good both at low and large
times. This is illustrated further in Fig.~\ref{h4_6} where we show the
time structure of muons arriving at transverse 
distances $r\sim $ 400~m, 1~km, and 2.5~km
compared to the results obtained through direct simulation.  These
results have been obtained using $\kappa =0.8$ as suggested by theory,
and $\lambda=1$, $\gamma=2.6$, $Q=0.17~$GeV/c and $\rho k$=0.2~GeV
km$^{-1}$. Our results are encouraging and we can state that
the main contributions to the time structure of muons in air showers
have been identified and understood.

\begin{figure}[htb]
\begin{center}
\includegraphics[width=10 cm]{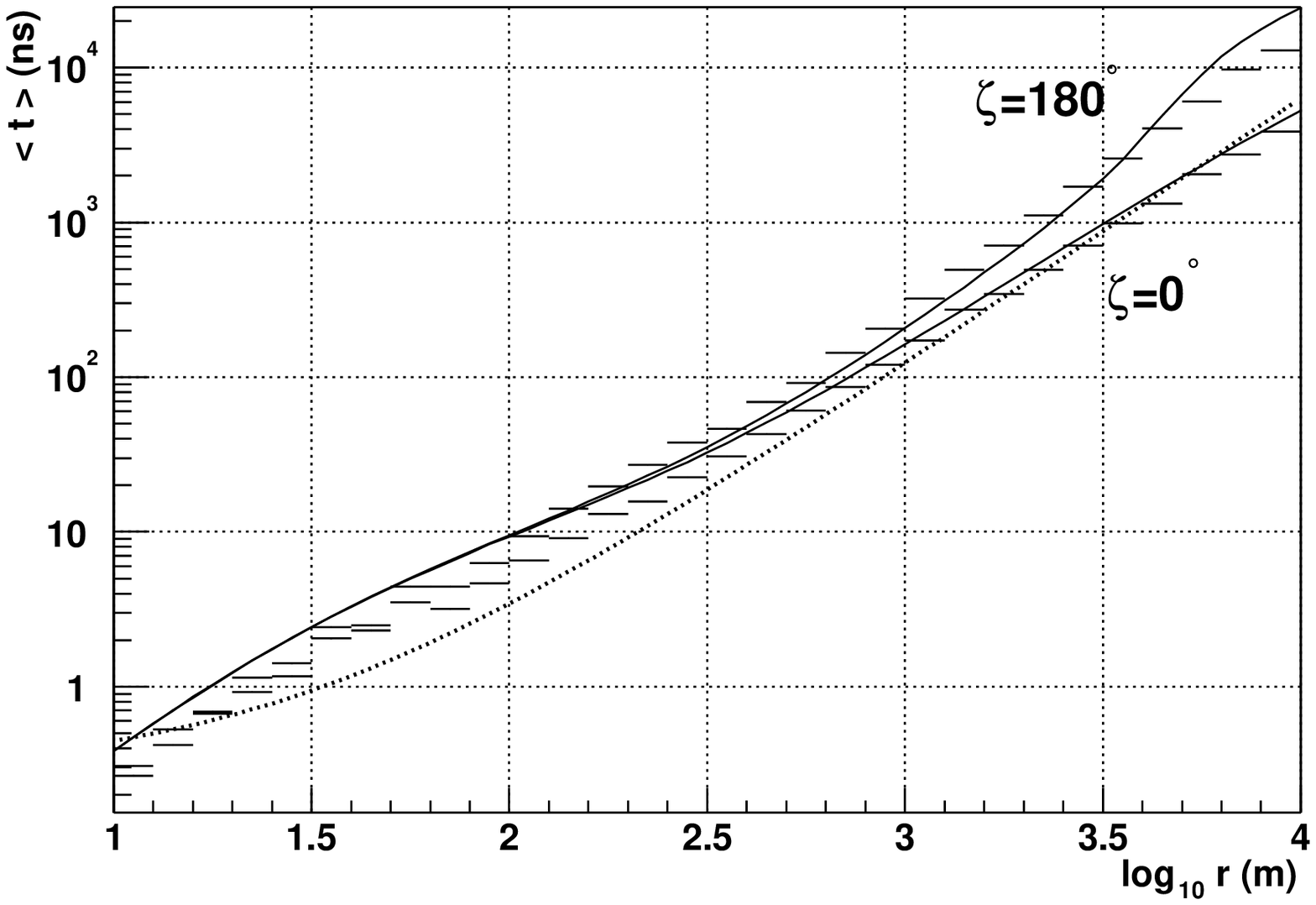}
\caption{Average time delay as a function of transverse distance 
for two different positions in the transverse plane corresponding to 
$\zeta = 0^\circ$ and $180^\circ$, as marked, for 10$^{19}$ eV proton 
showers at zenith angle 60$^\circ$. 
The histograms show the simulation results using AIRES and the lines show 
our analytical calculation. The lower 
(dotted)line shows the result of the phenomenological parameterization 
given in \cite{time1}.}
\label{ave_60}
\end{center} 
\end{figure}
\begin{figure}[htb]
\begin{center}
\includegraphics[width=10 cm]{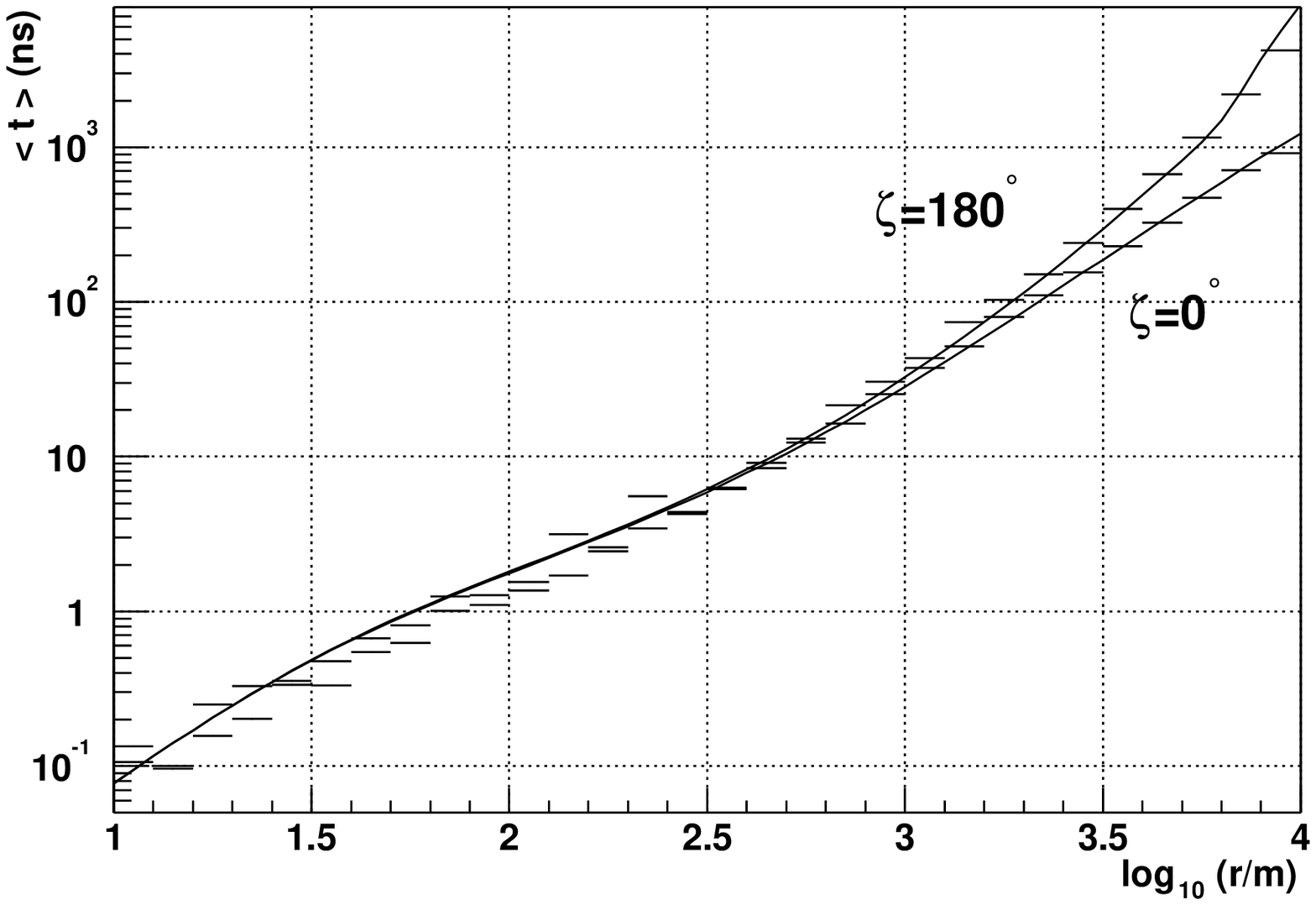}
\caption{Same as Fig. \ref{ave_60} for $80^\circ$.}
\label{ave_80}
\end{center} 
\end{figure}
\begin{figure}[htb]
\begin{center}
\includegraphics[width=10 cm]{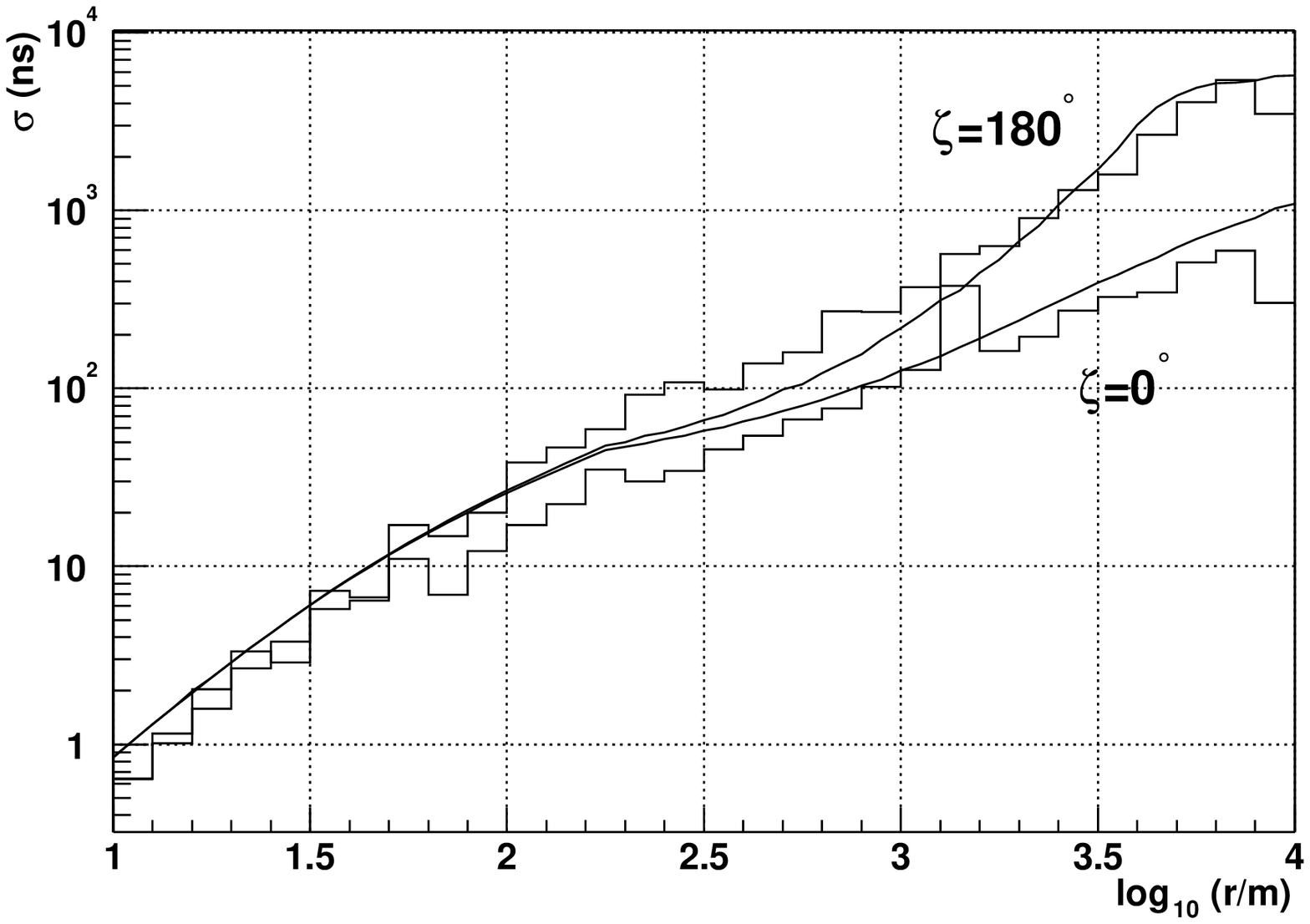}
\caption{RMS values for the arrival time distributions of muons as a 
function of transverse distance for $10^{19}$ eV proton showers at 
zenith angle $60^\circ$ and for two different angles $\zeta$. 
The histograms show the simulation results using AIRES and the  
lines show our analytical calculation.}
\label{rms_60}
\end{center} 
\end{figure}
\begin{figure}[htb]
\begin{center}
\includegraphics[width=10 cm]{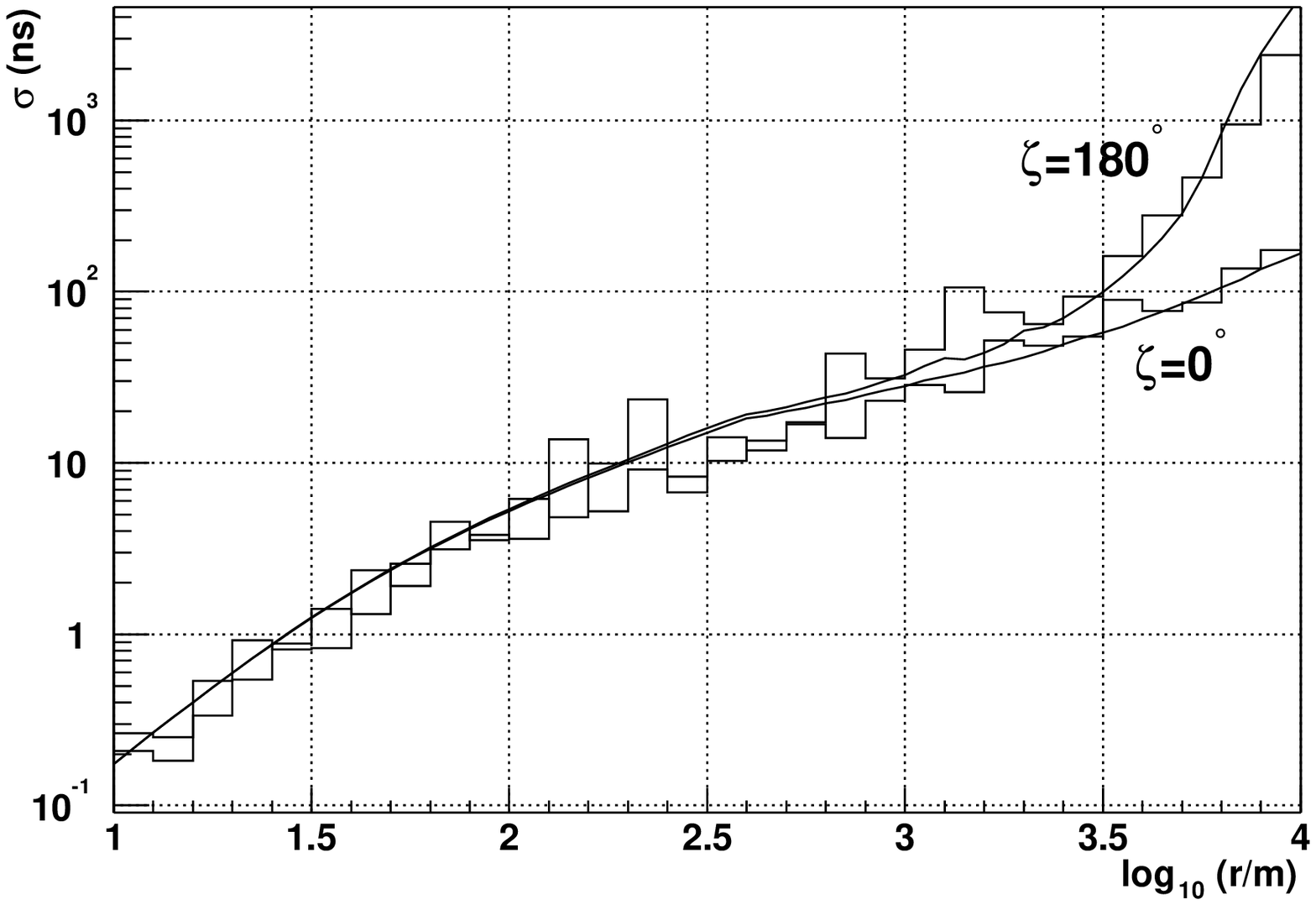}
\caption{Same as Fig.~\ref{rms_60} for $80^\circ$.}
\label{rms_80}
\end{center} 
\end{figure}

As an application of practical importance for direction reconstruction
in extensive air showers we have calculated the average and the width
of the arrival delay time distributions for different zenith angles as
a function of distance to the shower axis, $r$.  For near-vertical
showers the production distance distribution can not be approximated
by a delta function, as discussed at the end of Section II.  The final
kinematical delay distribution has been obtained by integrating the time
delay corresponding to a given production distance $z$ over the $z$
distribution
\begin{equation}
\varepsilon(t;r)=\int h(z) \varepsilon(t;z,r) dz.
\end{equation}
In Figs.~\ref{ave_60}, \ref{ave_80}, \ref{rms_60}, \ref{rms_80} 
we show the average and RMS values of the time distribution for times 
obtained with AIRES for 10$^{19}$ eV proton showers
and our analytical calculation for 60$^\circ$ and 80$^\circ$.
It can be seen that the agreement is 
good specially the larger the zenith angle. The magnitude of the 
asymmetry is well
reproduced as can be seen from the figure. For the RMS there are some
discrepancies both at large and small distances. We believe that this
is mainly an effect of fluctuations due to shower development and also
due to the thinning procedure adapted in AIRES. In Fig.~\ref{ave_60} we also
show the parameterization given in Ref.~\cite{time1}, based on a
phenomenological formula suggested by Linsley \cite{Linsley}.
\begin{figure}[htb]
\begin{center}
\includegraphics[width=10 cm]{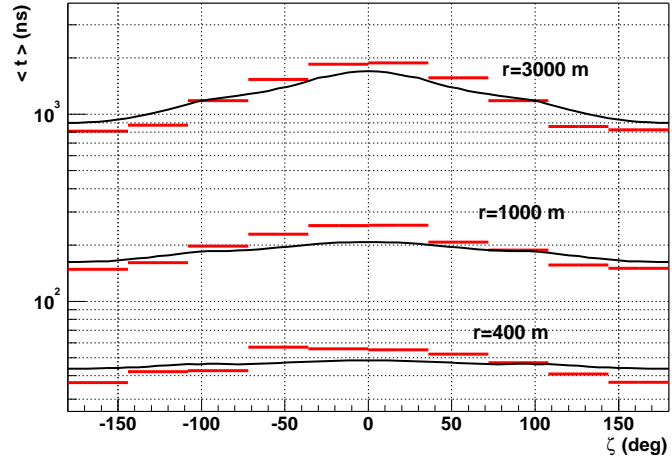}
\caption{Asymmetry in the average time delay 
which is plotted as a function of the angle $\zeta$ in the
transverse plane for 10$^{19}$ eV proton showers at 60$^\circ$ and fixed $r$ 
as marked. The histograms show the simulation 
results using AIRES and the lines show our analytical calculation.}
\label{dipole_60}
\end{center} 
\end{figure}
\begin{figure}[htb]
\begin{center}
\includegraphics[width=10 cm]{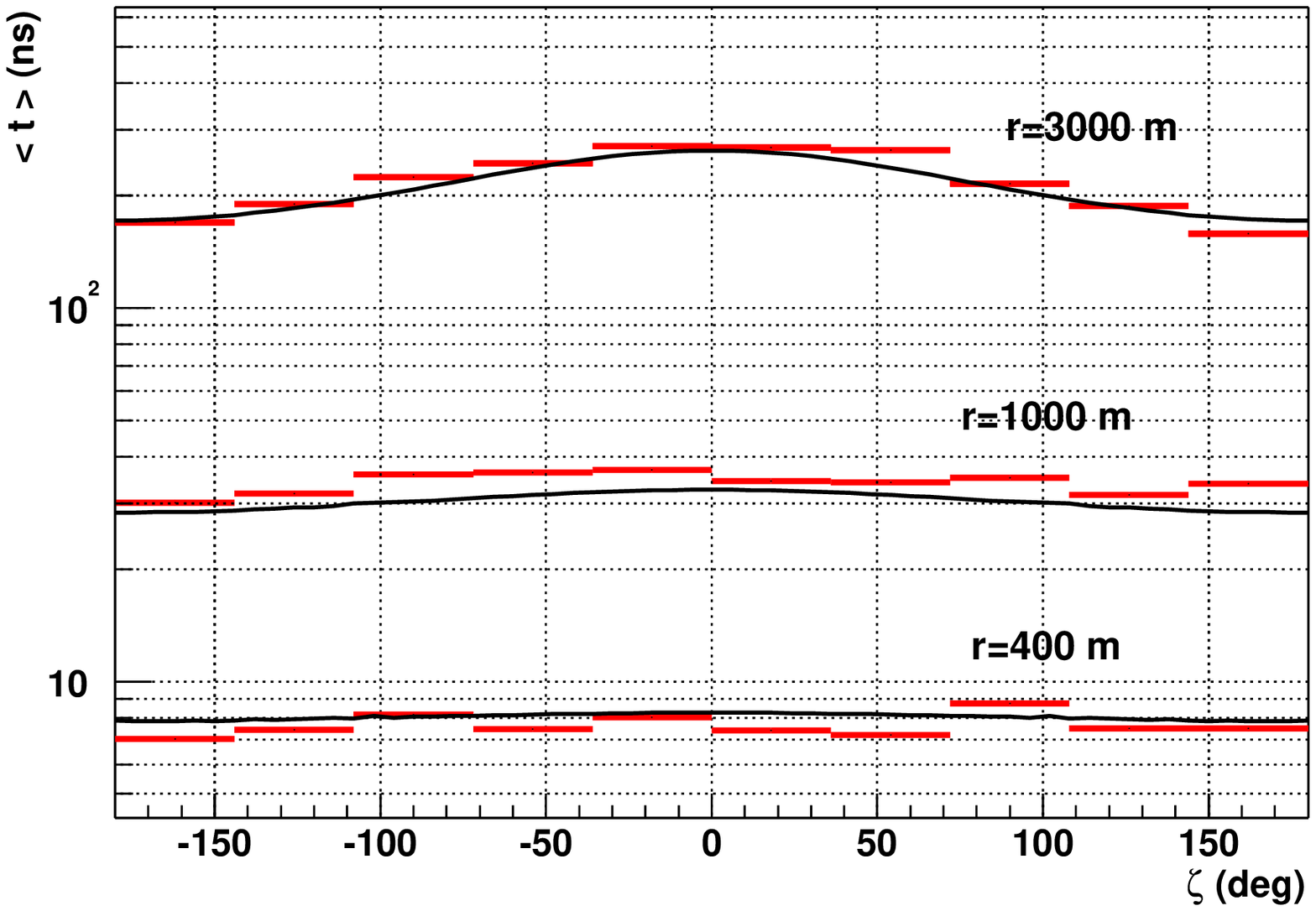}
\caption{Same as Fig.~\ref{dipole_60} for zenith angle of $80^\circ$.}
\label{dipole_80}
\end{center} 
\end{figure}

We have also calculated the asymmetries in the arrival time
distributions.  These can be approximated by a dipole correction to
the average over $\zeta$ in the transverse plane. 
In Figs.~\ref{dipole_60}, \ref{dipole_80} we give the magnitude of the
asymmetry as a function of $r$ for different zenith. The comparison
between the analytical and the 
simulation results are given in a separate graph for characteristic
distances.

We consider that an important application of this method will be to 
provide a description of the time distributions of muons in a 
given detector. Since
the distributions used are continuous, at some point the number of
muons that reach a given detector will have to be selected, and 
a sampling effect will arise because the finite number of
muons that pass through a given detector. This sampling effect
must be considered using the muon number distributions at ground level
which have been discussed in detail in ref.~\cite{ModelPaper}.

Up to now the magnetic field has been ignored but it will affect
the time structure of the signal in two ways. Firstly muon delays are
modified because particles do not travel in straight lines.  Secondly
significant modifications of the time delays calculated so far are to
be expected because the simple geometrical relation between $r$,
$E_i$, $z$ and $p_t$ is known not to hold. The effect of the magnetic
field on this relation is the basis of the model in
ref.~\cite{ModelPaper} used to describe de density profiles of muons in
inclined showers.  There it is shown that the relative importance of
the muon magnetic deflections with respect to deviations due to
their $p_t$ is given by the parameter $\alpha=z B/p_t$. The condition
of $\alpha > 1$ corresponds to $\theta>80^\circ$ at the Auger
site. Clearly when $\alpha \ll 1$ one can expect our description to be
valid. We have also tested the validity of our description for angles
corresponding to $\alpha > 1$ at the Auger site.  Contrary to what
could be expected, magnetic effects do not appear to modify the shape of
the time distributions in an important way for zenith angles below
$84^\circ$.  Therefore the results obtained in this article without
the magnetic field are phenomenologically interesting also in the
region $\alpha \sim 1$ and for the Auger site the range of application
is zenith angles below $\sim 84^\circ$.

The model described here on its own has also ignored shower
fluctuations.  Nevertheless fluctuations of the arrival times of muons
in an air shower can be related through the work presented here to
fluctuations in the distributions discussed here. It is reasonable to
assume that time fluctuations will be mostly due to the fluctuations
in the longitudinal
development of the shower and in the energy distribution of muons arriving 
at a given ground area.

\section{Summary and Conclusions}

We have obtained a procedure that can be used to describe the time
structure of the muon signal in extensive air showers. This procedure
is based on two simple mechanisms, time delays due to the extra path
length which are discussed as geometrical delays and those due to the
sub-luminal velocity of the muons. The global delay can be expressed as
a simple convolution of these two delays. The description combines the
physical mechanisms with phenomenological descriptions of the energy,
transverse momentum and position of the muons that are produced in an
air shower with several approximations. The results obtained agree
sufficiently well with those obtained with direct simulations for most
distances of interest and for all zenith angles to be of a great
importance for practical applications. This agreement allows us to state
with some confidence that these two mechanisms dominate the time
structure of the muons in air showers.

\section*{Acknowledgments} 

LC and 
AAW acknowledge the hospitality of the Center for Cosmological Physics, 
University of Chicago, where this work was initiated. 
We also thank Maximo Ave and James Cronin for many discussions. 
This work was partly supported by the Xunta de Galicia 
(PGIDIT02 PXIC 20611PN), by MCYT (FPA 2001-3837 and FPA 2002-01161). 
RAV is supported by the ``Ram\'on y Cajal'' program. 
We thank CESGA, ``Centro de Supercomputaci\'on de Galicia'' for computer 
resources. We thank the ''Fondo social europeo'' for support.
AAW acknowledges continuing support from PPARC (UK).


\begin{thebibliography}{999}
\bibitem{ICRC}  See Auger papers to be presented at the ICRC
  conference in Tokyo.
\end{thebibliography}
\end{document}